\documentclass{ws-procs9x6}            

\usepackage{xspace}

\newcommand{\pb}{\mbox{Pb--Pb}\xspace}

\begin{document}
\title{Light flavor baryon production from small to large collision systems at ALICE}
\author{D. Colella for the ALICE Collaboration}
\address{Istituto Nazionale di Fisica Nucleare,\\
via E. Orabona 4 \\
Bari, 70125, Italy\\
domenico.colella@cern.ch}

\begin{abstract}
Studies of light hadron and nuclei production are fundamental to characterize the hot and dense 
fireball created in ultra-relativistic heavy ion collisions and to investigate hadronisation 
mechanisms at the LHC. Observables investigated as a function of the charged particle multiplicity 
in proton-proton and proton-lead collisions have shown features not expected and qualitatively 
similar to what has been observed in larger size colliding systems. The ALICE experiment, exploiting 
its excellent tracking and PID capabilities, has performed an extensive and systematic study of 
strange and non-strange hadrons, short-lived hadron resonances and light (anti-)(hyper-)nuclei. 
A critical overview of these results will be presented through comparison with the statistical 
hadronisation model.
\end{abstract}

\bodymatter
\section{Introduction}
To the best of our knowledge, hot and dense matter
is produced in heavy-ion collisions at the Large Hadron Collider (LHC) energies. 
The system cools down and undergoes a transition to hadron gas. While 
the particle yields are fixed at the moment when the rate of inelastic collisions becomes negligible
(chemical freeze-out), the transverse momentum distributions continue to change until elastic
interactions cease (kinetic freeze-out). Studying hadrons and light nuclei containing light 
flavor and strange quarks is fundamental to better characterize the scenario described above. 
The yield of light flavor hadrons containing strange quarks is  
measured to study the dependency of the strangeness enhancement by event activity, colliding system,
colliding energy and strangeness content of the particles and to verify if it is present also for
particles with hidden strangeness.
Production yield of short-lived resonances with lifetimes comparable to that of the fireball is measured
to study rescattering and regeneration processes in the dense hadronic medium. 
The yield of nuclei and anti-nuclei in heavy-ion collisions provide information about the 
chemical freeze-out temperature and it is very intriguing to understand how such loosely bound
systems can survive the hot fireball. 

The ALICE Collaboration performed a systematic study of non-strange hadrons
($\pi^{+}$, $\mathrm{K}^{+}$, $\rm{p}$), single and multi-strange weak decaying hadrons 
($\mathrm{K^{0}_{S}}$, $\Lambda$, $\Xi^{-}$, $\Omega^{-}$), short lived resonances 
($\rho(770)^{0}$, $\mathrm{K^{*}(892)^{0}}$, $\phi(1020)$, $\Sigma^{*}(1385)^{+}$, 
$\Lambda(1520)$, $\Xi(1530)^{0}$), light nuclei ($\mathrm d$, $\mathrm t$, 
$^{3}$He and $^{3}_{\Lambda}$H) and the corresponding antiparticles, in many colliding systems at 
various energies provided by the LHC\cite{A,B,C,D,E,F,G,H,I,J,K}. 
The resonances will be referred to only by their commonly used symbol.

\section{Results and observations}
\begin{figure}
\centering
\includegraphics[width=27pc,clip]{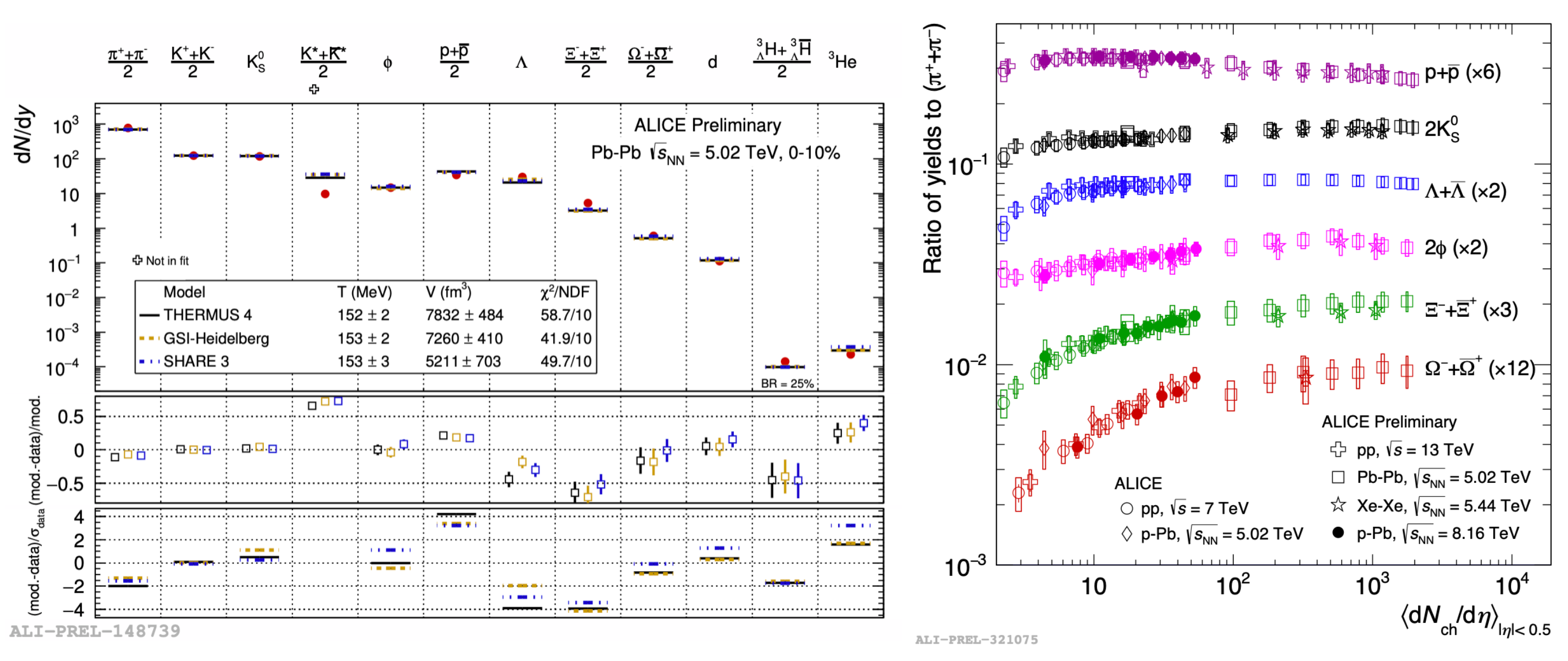}
\caption{(left) Particle yields compared to three grand-canonical SHM predictions in 0-10\% 
central \pb collisions at $\sqrt{s_{\rm NN}}$ = 5.02 $\mathrm{TeV}$. (right) Ratios of particle 
yields measured as a function of multiplicity across different collision systems.}
\label{fig-1}
\end{figure}

Particle yields in 0-10\% central \pb collisions at $\sqrt{s_{\rm NN}}$ = 5.02 $\mathrm{TeV}$ are 
compared to predictions from three Statistical Hadronization Models (SHMs), each based on a 
grand-canonical 
ensemble, and are shown in the left panel of Figure \ref{fig-1}.
The models assume hadron production at chemical equilibrium and reproduce most of the measured 
yields within uncertainties. The estimated chemical freeze-out temperature is about 153 $\mathrm{MeV}$, 
same as the value obtained in describing the data in \pb collisions at \mbox{$\sqrt{s_{\rm NN}}$ = 2.76 
$\mathrm{TeV}$}\cite{K,L}. 
A remarkable exception is the $\mathrm{K^{*0}}$, for which the deficit with respect to the
predicted yield can be explained by loss of $\mathrm{K^{*0}}$ signal in the hadronic phase.
There is also a tension for protons and multi-strange baryons whose explanation requires
additional effects (baryon annihilation, interactions in the hadron gas or feed-down from excited 
hadronic states, etc.)\cite{F,C}.
It is notable that the models also reproduce production of nuclei and hyper-nuclei although their 
binding energies are much smaller than the extracted values of the chemical freeze-out temperature. 
Overall, SHMs are quite successful and describe yields of production of hadrons which vary over 
seven orders of magnitude.

One of the most remarkable results established at the LHC is the smooth evolution of particle
chemistry from small to large systems as a function of charged particle multiplicity\cite{A}.
The particle yields normalized to charged-pion yield evolve monotonically and continuously as 
a function of charged particle multiplicity independent of collision energy and system size, 
as shown in Figure \ref{fig-1} (right).
These data also demonstrate, for the first time, the presence of strangeness enhancement 
in high \mbox{multiplicity} collisions of small systems, the effect being stronger for particles 
with higher strangeness content. The original interpretation of this phenomenon, considered as a 
signature of QGP formation in heavy-ion collisions, is not anymore straightforward.
In the strangeness canonical approach, the multiplicity dependence can be explained by requiring
local strangeness conservation while the bulk of the particles can still be described in the 
grand-canonical ensemble. 
Deviation from this description is present for the $\mathrm{K^{*0}}$, 
while a better description is obtained for $\Xi$ and $\Omega$
applying core-corona corrections. An interesting deviation, is also present for the $\phi$ meson:
being a strangeness-neutral particle, a flat multiplicity dependence is predicted, but a trend
is observed.  
   
\begin{figure}[t]
\centering
\includegraphics[width=27pc,clip]{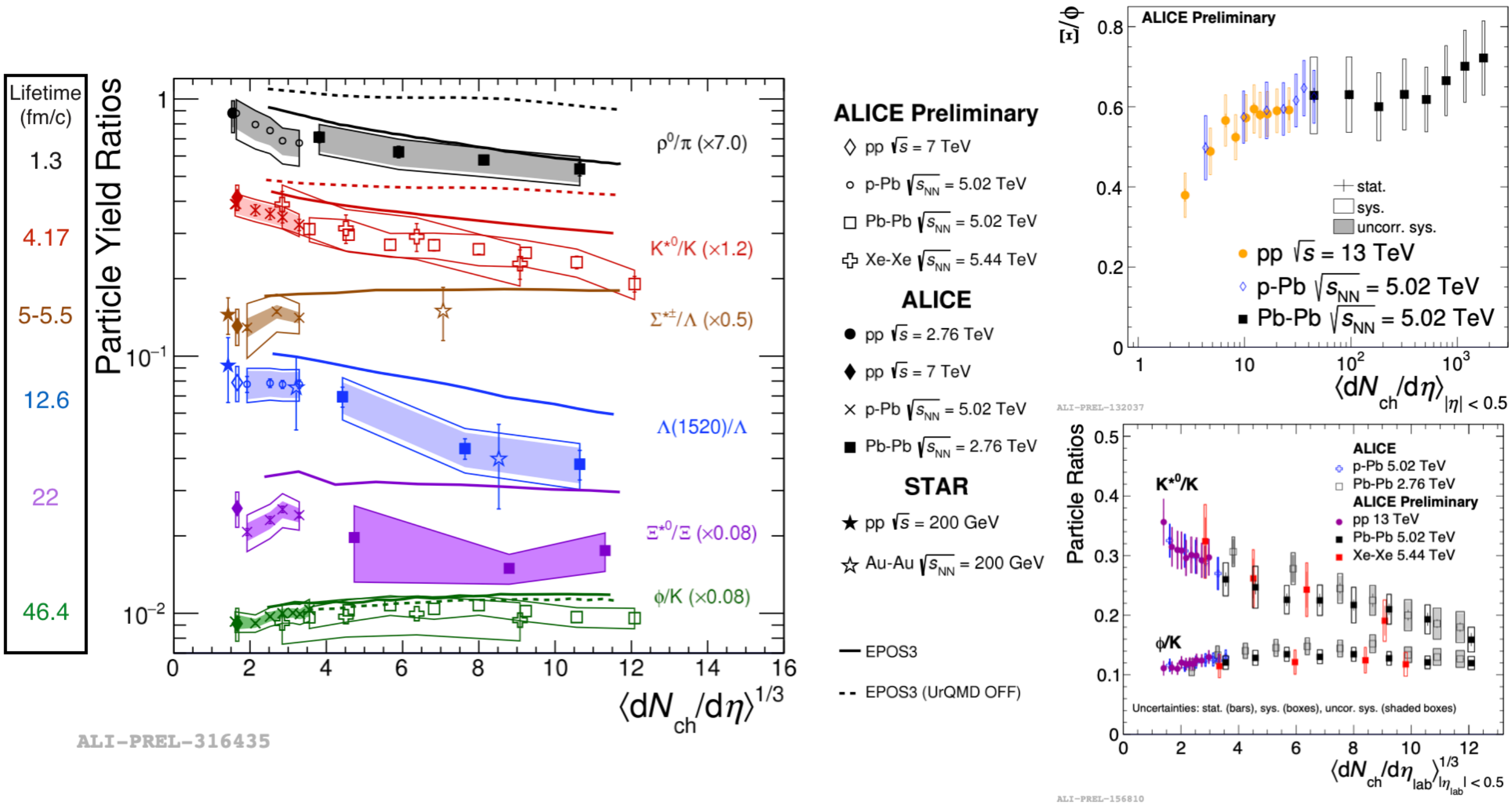}
\caption{(left) $\rho^{0}$/$\pi$, $\mathrm{K^{*0}}$/K, 
$\Sigma^{\pm}$/$\Lambda$, $\Lambda(1520)$/$\Lambda$, $\Xi^{0}$/$\Xi$ and
$\phi$/K yield ratios as a function of multiplicity. (right) $\Xi$/$\phi$ 
and $\phi$/K yield ratios as a function of multiplicity.}
\label{fig-2}
\end{figure}

Effect of particle rescattering and regeneration in the hadronic phase can be probed by measuring
the yield of hadronic states having a lifetime comparable to the one of the fireball. In this
picture the resonance yields in the final state are determined by the resonance yields at chemical
freeze-out, their lifetimes, hadronic phase lifetime and by scattering cross sections.
Figure \ref{fig-2} (left) shows ratios of the resonance yields to yields of stable hadrons having similar
constituent quarks, for a set of particles covering a wide range of lifetimes from 1.3 fm/$\it{c}$
for the $\rho^{0}$ meson up to 46.4 fm/$\it{c}$ for the $\phi$ meson. 
The ratios are suppressed in central heavy-ion collisions for resonances
with lifetimes $\tau <$ 20 fm/$\it{c}$ compared to pp and peripheral heavy-ion collisions.
For longer-lived resonances, as the $\phi$, ratios are not suppressed. 
These results support the existence of a hadronic phase that lives long enough
to cause a significant reduction of the reconstructed yields of short lived resonances.
Resonance measurements allow for estimates of the hadronic phase lifetime: 10 fm/$\it{c}$ for the 
most central \pb collisions and 1-2 fm/$\it{c}$ for the most peripheral \pb collisions.
These values are estimated from the EPOS+UrQMD model\cite{M}.

In order to investigate the behavior of hidden strangeness, the following two ratios have been
measured as a function of charged particle multiplicity and are shown in 
Figure \ref{fig-2} (right): $\Xi$($|S|$=2)/$\phi$($|S|$=0) and $\phi$($|S|$=0)/K($|S|$=1). 
The multiplicity
evolution of these two ratios suggests that the $\phi$ meson behaves as if it had between
1 and 2 units of strangeness, so that the $\Xi$ is more enhanced than $\phi$, which is more
enhanced than $K$.  

\begin{figure}[t]
\centering
\includegraphics[width=20pc,clip]{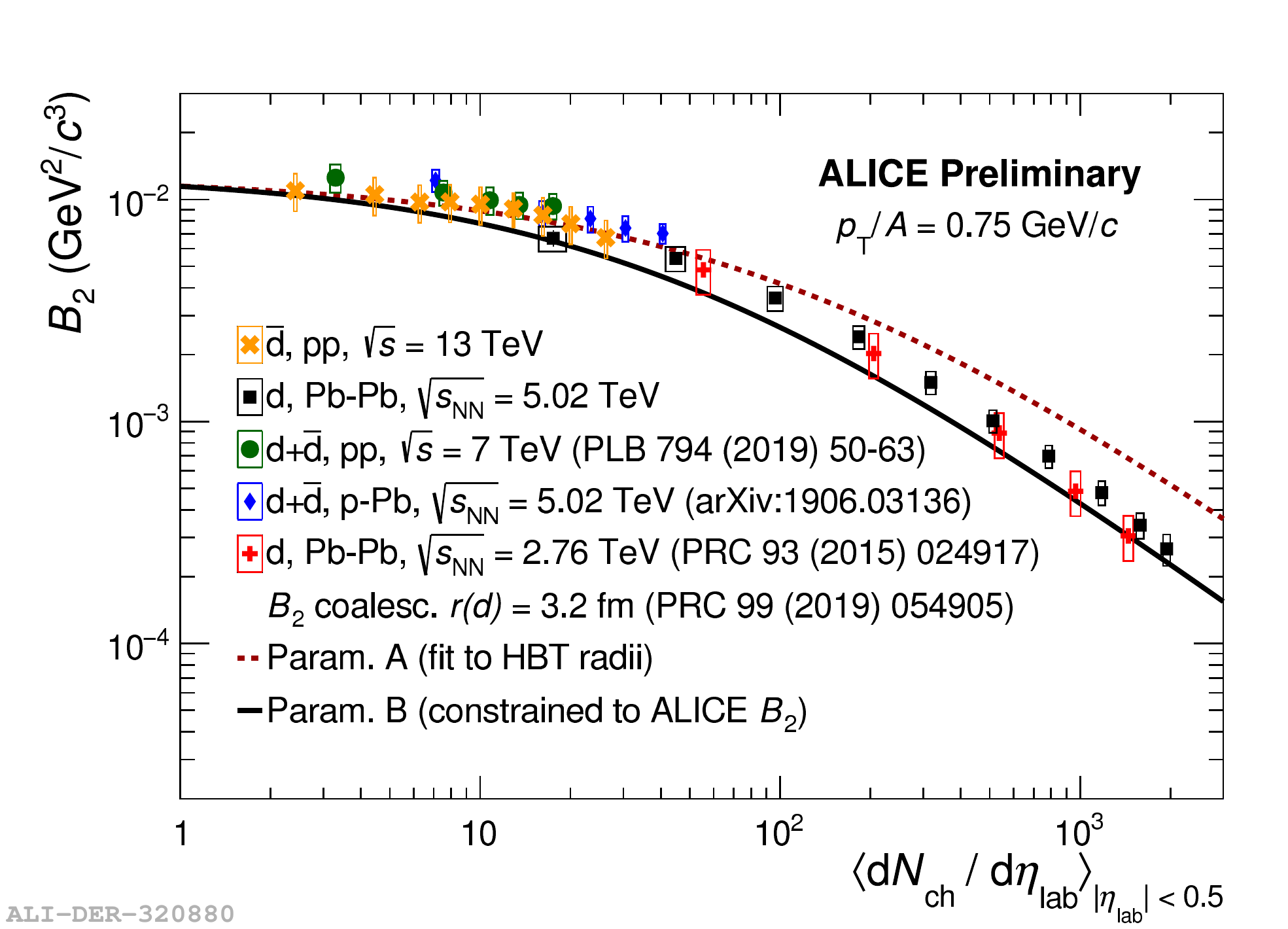}
\caption{Anti-deuteron coalescence parameter $\it{B}_{2}$ at fixed p$_{T}$/$\it{A}$ = 0.75 
GeV/$\it{c}$ as a function of the average charged-particle multiplicity.}
\label{fig-3}
\end{figure}

Light nuclei are characterised by a low binding energy (EB $\sim$ 1 MeV) compared to the 
temperature of the chemical freeze-out, (T$_{ch}$ $\sim$ 160 MeV). Therefore, in principle 
one would not expect to observe any nucleus. The experimental results on their production
are usually interpreted within two approaches. According to the SHM, (anti-)nuclei are
produced at the chemical freeze-out in statistical equilibrium, along with all the other 
hadrons. As already shown this model nicely predict the yields for $\mathrm d$, $^{3}_{\Lambda}$H
and $^{3}$He in \mbox{0-10\%} central \pb collisions. In the coalescence picture, nucleons that 
are close to each other in phase space after chemical freeze-out can merge and form a nucleus 
via coalescence. In Figure \ref{fig-3} the coalescence parameter for the deuterons ($B_{2}$)
is shown as a function of charged particle multiplicity in many colliding systems\cite{B,J}. This
is the main observable of the model, related to the probability to form a nucleus via
coalescence. $B_{2}$ is independent of the collision system or the energy, 
and depends only on the charged particle density. This suggest a common production mechanism 
that depends only on the system size.
In the current state more data and more precise model calculations are needed to discern the
production mechanisms of light nuclei.

\end{document}